\documentclass[letterpaper,twocolumn,10pt]{article}
\usepackage{usenix}

\usepackage{amsmath}
\usepackage{xcolor}
\usepackage{xspace}
\usepackage{graphicx}
\usepackage{booktabs}
\usepackage{tabularx}
\usepackage{makecell}
\usepackage{authblk}
\usepackage{rotating}
\usepackage{tablefootnote}
\usepackage{listings}
\usepackage{hyperref}

\newcommand{\cplat}{CPLAT\xspace}
\newcommand{\hcs}{container shim\xspace}
\newcommand{\gcs}{guest agent\xspace}
\newcommand{\uvm}{UVM\xspace}
\newcommand{\system}{Parma\xspace}
\newcommand{\nginx}{\texttt{nginx}\xspace}
\newcommand{\containerd}{\texttt{containerd}\xspace}
\newcommand{\redis}{\texttt{redis}\xspace}
\newcommand{\eg}{\textit{e.g.,}\xspace}
\newcommand{\ie}{\textit{i.e.,}\xspace}
\newcommand{\etal}{\textit{et al.}\xspace}

\lstdefinestyle{mystyle}{
    numberstyle=\tiny,
    basicstyle=\ttfamily\footnotesize,
    breakatwhitespace=false,         
    breaklines=true,                 
    captionpos=b,                    
    keepspaces=true,                 
    numbers=left,                    
    numbersep=5pt,                  
    showspaces=false,                
    showstringspaces=false,
    showtabs=false,                  
    tabsize=2
}

\begin{document}
%-------------------------------------------------------------------------------

%don't want date printed
\date{}
\title{\Large \bf \system: Confidential Containers via Attested Execution Policies}

\author{Matthew A. Johnson}
\author{Stavros Volos}
\author{Ken Gordon}
\author{Sean T. Allen}
\author{Christoph M. Wintersteiger}
\author{Sylvan Clebsch}
\author{John Starks}
\author{Manuel Costa}
\affil{Azure Research}

\maketitle

%-------------------------------------------------------------------------------
\begin{abstract}
    %-------------------------------------------------------------------------------
    Container-based technologies empower cloud tenants to develop highly portable software and deploy services in the cloud at a rapid pace.
    Cloud privacy, meanwhile, is important as a large number of container deployments operate on privacy-sensitive data, but challenging due to the increasing frequency and sophistication of attacks.
    State-of-the-art confidential container-based designs leverage process-based trusted execution environments (TEEs), but face security and compatibility issues that limits their practical deployment.

    We propose \system, an architecture that provides lift-and-shift deployment of unmodified containers while providing strong security protection against a powerful attacker who controls the untrusted host and hypervisor.
    \system leverages VM-level isolation to execute a container group within a unique VM-based TEE.
    Besides container integrity and user data confidentiality and integrity, \system also offers container attestation and execution integrity based on an attested execution policy.
    \system execution policies provide an inductive proof over all future states of the container group.
    This proof, which is established during initialization, forms a root of trust that can be used for secure operations within the container group without requiring any modifications of the containerized workflow itself (aside from the inclusion of the execution policy.)

    We evaluate \system on AMD SEV-SNP processors by running a diverse set of workloads demonstrating that workflows exhibit 0--26\% additional overhead in performance over running outside the enclave, with a mean 13\% overhead on SPEC2017, while requiring no modifications to their program code. Adding execution policies introduces less than 1\% additional overhead.
    Furthermore, we have deployed \system as the underlying technology driving Confidential Containers on Azure Container Instances.
\end{abstract}

%-------------------------------------------------------------------------------
\section{Introduction}
%-------------------------------------------------------------------------------

\begin{figure*}[t]
      \centering
      \begin{minipage}{0.48\textwidth}
            \centering
            \includegraphics[width=\linewidth]{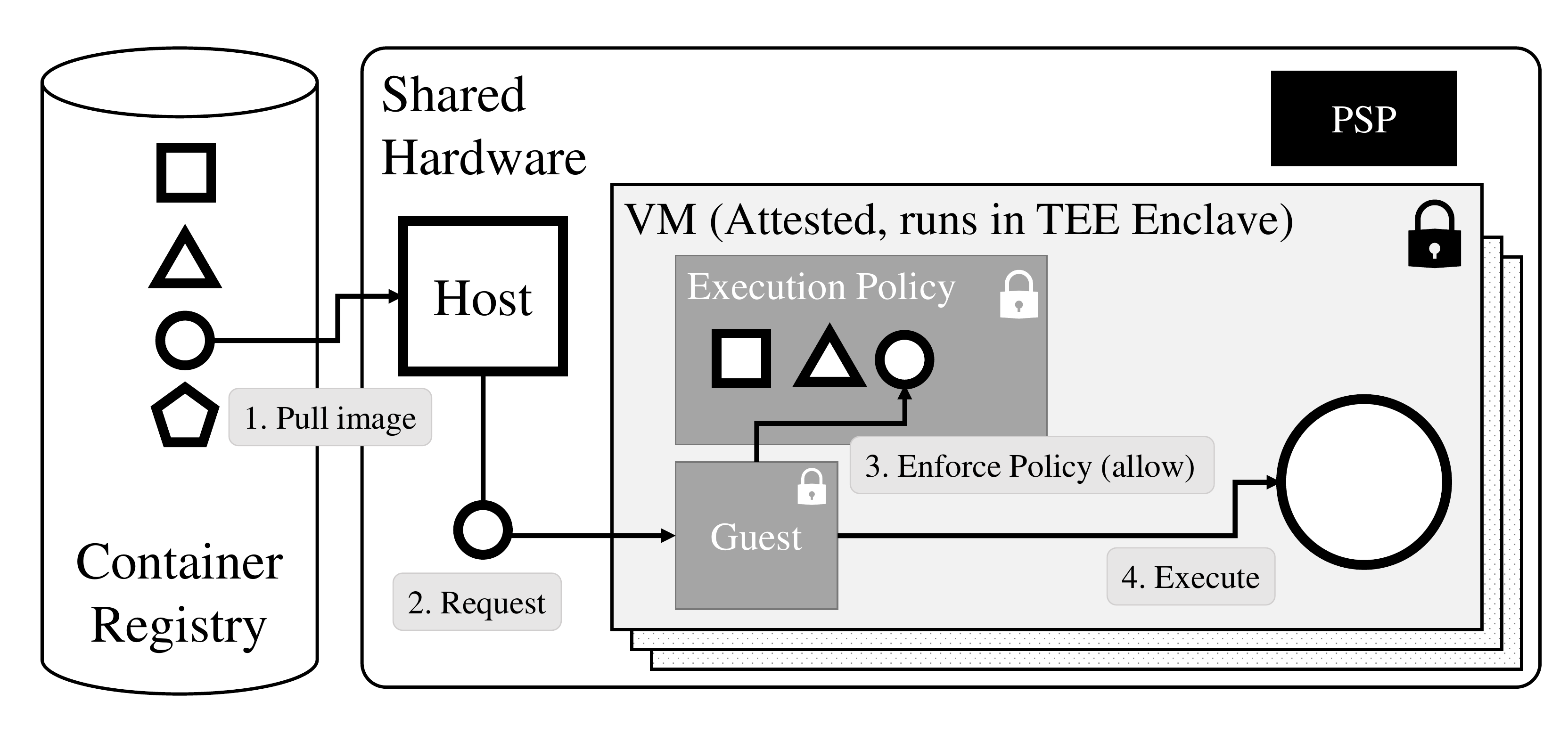} \\
            (a)
      \end{minipage}
      \begin{minipage}{0.48\textwidth}
            \centering
            \includegraphics[width=\linewidth]{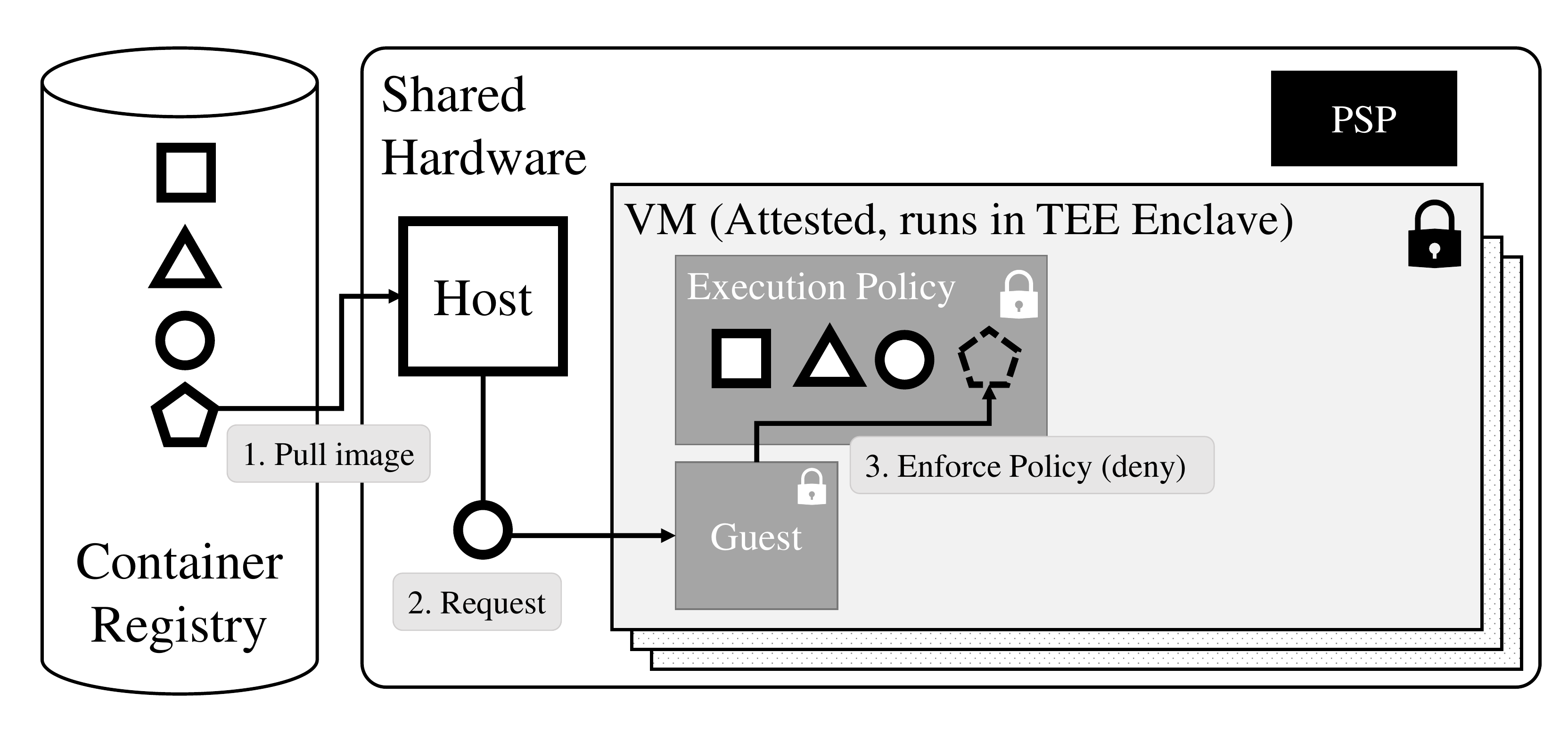} \\
            (b)
      \end{minipage}
      \caption{
            \textbf{Execution Policy}. The execution policy is a component of the utility VM that is attested at initialization time.
            It describes all of the actions the user has explicitly allowed  the \gcs to take within the container group.
            In (a) we see an example of a successful mount action, in which a layer of a container image has a dm-verity root hash which matches a hash enumerated in the policy.
            When the hash does not match, as in (b), this action is denied.}
      \label{fig-execution-policy}
\end{figure*}

Since the launch of the large-scale Infrastructure-as-a-Service (IaaS) offerings from Amazon (AWS in 2006), Microsoft (Azure in 2008), and Google (GCP in 2008), there has been a continuous trend towards cloud computing, which allows customers to leverage capability and cost advantages through economies of scale.
This was made possible through virtualization \cite{barham2003xen}, whereby virtual machines (VMs) allow the efficient use of large bare-metal compute architectures (hosts) by using a hypervisor to coordinate sharing between multiple tenants according to their expressed usage requirements.
However, while VMs provide a way for users to quickly obtain additional compute capacity and maximize the utilization of existing hardware (and/or avoid the cost of maintaining peak capacity by utilizing a public cloud), it is still necessary to configure, deploy, manage, and maintain VMs using traditional techniques.

In recent years, container-based technologies, such as Docker \cite{docker} and Kubernetes \cite{k8s} have arisen to address this orthogonal need, providing a lightweight solution for creating a set of machine configurations, called containers, which can be deployed onto hardware (virtualized or physical) as a group via an automated process.
Container technology provides multiple separated user-space instances which are isolated from one another via kernel software.
Unlike VMs, containers run directly on the host system (sharing its kernel) and as such do not need to emulate devices or maintain large disk files.
Further, containers defined according to the OCI Distribution Specification \cite{oci} specify dependencies as \emph{layers} which can be shared between different containers, making them amenable to caching and thus speeding up deployment while reducing storage costs for multiple containers.
The success of containerization technology for on-premises systems has led to major cloud providers developing their own Container-as-a-Service (CaaS) offerings \cite{aci,amazon_docker,google_cloud_run} which provide customers the ability to maintain and deploy containers in the public cloud.
In CaaS offerings, containers run in a per-group utility VM (\uvm) which provides hypervisor-level isolation between containers running from different tenants on the same host.
While the container manager and \hcs running on the host are responsible for pulling images from the container registry, bringing up the utility VM, and orchestrating container execution, an agent running within the utility VM (the \gcs) coordinates the container workflow as directed by the host-side \hcs.

Cloud computing poses unique risks.
Although VMs and VM-isolated container groups provide strong isolation between tenants, they are deployed by the cloud service provider (CSP) and coordinated by the CSP's hypervisor.
As such, the host operating system (including the container manager and \hcs) and the hypervisor all lie within the Trusted Computing Base (TCB).
Research into confidential cloud computing attempts to reduce the attack surface by leveraging specialized hardware-enforced Trusted Execution Environments (TEEs) \cite{bahmani20cure, brasser19sanctuary,champagne10scalable,costan16sanctum,evtyushkin14isox, lee20keystone,mccune08flicker,suh03aegis,sun15trustice}, which enable user workloads to be protected inside \emph{enclaves} even if the host's software is compromised or controlled by a malicious entity.
TEEs available from major CPU vendors can be either \textit{process-based}, such as Intel SGX \cite{intel_sgx} and ARM TrustZone \cite{arm_trustzone}, or \textit{VM-based}, such as AMD SEV-SNP \cite{amd_sev_snp, amd_sev}, Intel TDX \cite{intel_tdx} and ARM CCA \cite{arm_rme, arm21cca}.
VM-based TEEs offer hardware-level isolation of the VM, preventing the host operating system and the hypervisor from having access to the VM's memory and registers.

With CaaS, container execution is orchestrated by a host-side shim that communicates with the \gcs, which coordinates the activity of the container group within the \uvm.
The \uvm can be hardware-isolated within a TEE enclave, but the container images are controlled by the host, as is the order in which they are mounted, the container environment variables, the commands that are sent to the containers via the bridge between the \hcs and the \gcs, and so forth.
This means that a compromised host can overcome the hardware isolation of the VM by injecting malicious containers.
This risk of attack, be it from malicious or compromised employees of the CSP or external threats, limits the extent to which containerization can be used in the cloud for sensitive workloads in industries like finance and healthcare.

The naive solution to this problem is to run the \gcs and \hcs within the same VM-based TEE.
This removes the CSP from the TCB, but it also removes the CSP's ability to orchestrate and automate the container workflow.
In addition, the container owner is then in the TCB.
The container images are controlled by the container owner, as is the order in which they are mounted, the container environment variables, and the commands that are sent to the containers.
The end-user of the confidential container (\eg a customer of a bank, a patient providing data to a doctor) must trust that the container owner has and will run only the expected commands.
This also leaves image integrity and data confidentiality and integrity unsolved.

\paragraph{Our work.} We present \system, an architecture that implements the \textit{confidential containers} abstraction on a state-of-the-art \texttt{containerd} \cite{containerd} stack running on processors with VM-based TEE support (\ie AMD SEV-SNP processors).
\system provides a lift-and-shift experience and the ability to run unmodified containers pulled from (unmodified) container registries while providing strong security guarantees:
\textit{container attestation and integrity}, meaning that only customer-specified containers can run within the TCB and any means of container tampering is detected by the TCB;
and \textit{user data confidentiality and integrity}, meaning that only the TCB has access to the user's data and any means of data tampering is detected by the TCB.

\system provides strong protection for the container's root filesystem (comprised of the container image layers and writeable scratch space) and the user's data.
For container image layers (pulled in plaintext by the untrusted container manager and stored in a host-side block device), \system mounts the device as an integrity-protected read-only filesystem (using dm-verity) and relies on the filesystem driver to enforce integrity checking upon an access to the filesystem.
For confidentiality and integrity of privacy-sensitive data stored in a block device (\eg writeable scratch space of the container's root filesystem) or blob storage (\eg remote blobs holding user data), \system relies on block-level encryption and integrity (using dm-crypt + dm-integrity) to decrypt memory-mapped blocks, guaranteeing that data appears in plaintext only within the VM's hardware-protected memory.

Finally, \system provides container attestation rooted in a hardware-issued attestation by enforcing attested user-specified \emph{execution policies}.
We have augmented the \gcs to enforce the execution policy such that it only executes commands (submitted by the untrusted \hcs) which are explicitly allowed by the user, as seen in Figure \ref{fig-execution-policy}.
The policy is attested by encoding its measurement in the attestation report as an immutable field at \uvm initialization.
As a result of including the execution policy, the hardware-issued attestation forms an inductive proof over the future state of the container group.
The attestation can then be used downstream for operations needed by secure computation.
For example, remote verifiers may release keys (governing the user's encrypted data) to only those container groups which can present an attestation report encoding the expected execution policy and measurement of the utility VM.

\paragraph{Contributions:} The main contributions of our work are:
\begin{itemize}
      \item \system, a novel security architecture for confidential containerized workloads.
            \system establishes security guarantees rooted in an inductive proof over all future states of the container group provided by the introduction of an attested execution policy.
      \item an implementation of \system which forms the basis for Confidential Containers on Azure Container Instances \cite{aci_cc} and is publicly available on GitHub \footnote{\url{https://github.com/microsoft/hcsshim/tree/main/pkg/securitypolicy}}.
      \item neither requiring changes to existing containers, nor container image signing.
            Instead, the execution policy ensures that only the actions the user has explicitly expressed are allowed to take place within the container group, maintaining support for existing CaaS deployment practices.
      \item an evaluation of our implementation with standard benchmarks for computation, network, and database activity.
            We compare a base container system to containers running within a TEE enclave with and without \system.
            We demonstrate that \system introduces 0--26\% additional overhead in performance over running outside the enclave, with a mean 13\% overhead on SPEC2017, and that adding execution policies introduces less than 1\% additional overhead.
\end{itemize}

There were also significant implementation challenges, including SEV-SNP enablement in the hypervisor, bounce buffers for I/O, Linux enlightenment for SEV-SNP including attestation report fetching, and hardening the hypervisor interface.
These are not claimed as contributions.

%-------------------------------------------------------------------------------
\section{Background}
%-------------------------------------------------------------------------------

We will begin by introducing the technological dependencies of \system, namely Trusted Execution Environments (TEEs) and the AMD SEV-SNP architecture.

\subsection{Trusted Execution Environments}
There have been many proposed security architectures which aim to provide a TEE~\cite{amd_sev, bahmani20cure,brasser19sanctuary,costan16sanctum,intel_sgx, lee20keystone}.
The goal of a TEE is to isolate workloads from the host system in order to protect them against manipulation whilst running.
Most architectures provide secure environments, typically called \emph{enclaves}, which run in parallel with the underlying host operating system.
As such, the Trusted Computing Base (TCB) contains the required hardware which provides the needed security capabilities and the software to utilize it to maintain the guarantees of the TEE.
While each TEE architecture has its own idiosyncrasies, there are desirable qualities which increase their utility:
\begin{description}
      \item [Small TCB] Minimizing the TCB is essential to reduce the attack surface of the TEE.
      \item [Strong Isolation] The enclaves must be isolated from the host at all times, including the register state and memory.
      \item [Attestable State] The boot-up process and state of the TEE must be verifiable using attestation.
      \item [Minimal Overhead] High performance costs incurred by using the TEE greatly minimize utility.
      \item [Minimal Adoption Cost] While perhaps not a goal shared by all TEEs, greater utility is achieved if running code in the TEE does not require significant reworking of a workflow
            (\eg rewriting software to target an enclave-specific subset of a language, requiring custom tool-chains).
\end{description}

\subsection{AMD Secure Encrypted Virtualization-Secure Nested Paging}
The commercially available TEE offering from AMD is called Secure Encrypted Virtualization (SEV) \cite{amd_sev} and targets cloud servers.
As indicated in the name, it is focused on protecting Virtual Machines (VMs) from a malicious host or hypervisor.
We use a specialization of SEV called SEV-SNP \cite{amd_sev_snp} (Secure Nested Paging).
AMD SEV-SNP is available in AMD's EPYC Milan processors and extends the SEV and SEV-ES (Encrypted State) technologies, which offer isolation of a VM by providing encrypted memory and CPU register state.
AMD SEV-SNP adds memory integrity protection to ensure that a VM is able to read the most recent values written to an encrypted memory page.
In doing so, it provides protection against data replay, corruption, remapping- and aliasing-based attacks.

\subsubsection{Platform Security Processor}
The Platform Security Processor firmware (PSP) implements the security environment for hardware-isolated VMs.
The PSP provides a unique identity to the CPU by deriving the Versioned Chip Endorsement Key (VCEK) from chip-unique secrets and the current TCB version.
The PSP also provides ABI functions for managing the platform, the life-cycle of a guest VM, and data structures utilized by the PSP to maintain integrity of memory pages.

\subsubsection{Memory Encryption}
AMD Secure Memory Encryption (SME) \cite{amd_sev} is a general-purpose mechanism for main memory encryption that is flexible and integrated into the CPU architecture.
It is provided via dedicated hardware in the on-die memory controllers that provides an Advanced Encryption Standard (AES) engine.
This encrypts data when it is written to DRAM, and then decrypts it when read, providing protection against physical attacks on the memory bus and/or modules.
The key used by the AES engine is randomly generated on each system reset and is not visible to any process running on the CPU cores.
Instead, the key is managed entirely by the PSP.
Each VM has memory encrypted with its own key, and can choose which data memory pages they would like to be private.
Private memory is encrypted with a guest-specific key, whereas shared memory may be encrypted with a hypervisor key.

\subsubsection{Secure Nested Paging}
The memory encryption provided by AMD-SEV is necessary but not sufficient to protect against runtime manipulation.
In particular, it does not protect against \emph{integrity attacks} such as:
\begin{description}
      \item [Replay] The attacker writes a valid past block of data to a memory page.
            This is of particular concern if the attacker knows the unencrypted data.
      \item [Data Corruption] If the attacker can write to a page then even if it is encrypted they can write random bytes, corrupting the memory.
      \item [Memory Aliasing] A malicious hypervisor maps two or more guest pages to the same physical page, such that the guest corrupts its own memory.
      \item [Memory Re-Mapping] A malicious hypervisor can also map one guest page to multiple physical pages, so that the guest has an inconsistent view of memory where only a subset of the data it wrote appears.
\end{description}

\subsubsection{Reverse Map Table}
The relationship between guest pages and physical pages is maintained by a structure called a Reverse Map Table (RMP).
It is shared across the system and contains one entry for every 4k page of DRAM that may be used by VMs.
The purpose of the RMP is to track the owner for each page of memory, and control access to memory so that only the owner of the page can write it.
The RMP is used in conjunction with standard x86 page tables to enforce memory restrictions and page access rights.
When running in an AMD SEV-SNP VM, the RMP check is slightly more complex.
AMD-V 2-level paging (also called Nested Paging) is used to translate a Guest Virtual Address (GVA) to a Guest Physical Address (GPA), and then finally to a System Physical Address (SPA).
The SPA is used to index the RMP and the entry is checked \cite{amd_sev_snp}.

\subsubsection{Page Validation}
Each RMP entry contains the GPA at which a particular page of DRAM should be mapped.
While the nested page tables ensure that each GPA can only map to one SPA, the hypervisor may change these tables at any time.
Thus, inside each RMP entry is a Validated bit, which is automatically cleared to zero by the CPU when a new RMP entry is created for a guest.
Pages which have the validated bit cleared are not usable by the hypervisor or as a private guest page.
The guest can only use the page after it sets the Validated bit via a new instruction, PVALIDATE.
Only the guest is able to use PVALIDATE, and each guest VM can only validate its own memory.
If the guest VM only validates the memory corresponding to a GPA once, then
the injective mapping between GPAs and SPAs is guaranteed.

\subsubsection{Attestation}
\label{sec-attestation}
The PSP can issue hardware attestation reports capturing various security-related attributes, constructed or specified during initialization and runtime.
Among other information, the resulting attestation report contains the \textit{guest launch measurement}, the \textit{host data}, and the \textit{report data}.
The attestation report is signed by the VCEK.

\paragraph{Initialization.} During VM launch, the PSP initializes a cryptographic digest context used to construct the measurement of the guest.
The hypervisor can insert data into the the guest's memory address space at the granularity of a page, during which the cryptographic digest context is updated with the data, thereby binding the measurement of the guest with all operations that the hypervisor took on the guest's memory contents.
A special page is added by the hypervisor to the guest memory, which is populated by the PSP with an encryption key that establishes a secure communication channel between the PSP and the guest.
Once the VM launch completes, the PSP finalizes the cryptographic digest which is encoded as the \textit{guest launch measurement} in the attestation report. The hypervisor may provide 256-bits of arbitrary data to be encoded as \textit{host data} in the attestation report.

\paragraph{Runtime.} The PSP generates attestation reports on behalf of a guest VM.
The request and response are submitted via the secure channel established during the guest launch, ensuring that a malicious host cannot impersonate the guest VM.
Upon requesting a report, the guest may supply 512-bits of arbitrary data to be encoded in the report as \textit{report data}.

%-------------------------------------------------------------------------------
\section{\system Architecture}
%-------------------------------------------------------------------------------
In this section, we present \system, an architecture that implements the \textit{confidential containers} abstraction using attested execution policies.
We first describe the container platform which forms the basis of the \system design and implementation (\ref{sec-cplat}) and then provide a detailed description of the threat model (\ref{sec-threat-model}).

Finally, we present the security guarantees under the threat model and how \system provides these guarantees via a collection of design principles (\ref{sec-guarantees}).
The guiding principle of \system is to provide an inductive proof over the state of a container group, rooted in the attestation report produced by the PSP.
The standard components and lifecycle for the container platform (the \cplat) are largely unchanged, with the exception of the \gcs, whose actions become circumscribed by the execution policy (\ref{sec-policy}).
Thus constrained, the future state of the system can be rooted in the measurement performed during guest initialization.

\subsection{Container Platform}
\label{sec-cplat}

The container platform (the \cplat) is a group of components built around the capabilities of \containerd \cite{containerd}.
\containerd is a daemon which manages the complete container life-cycle, from image pull and storage to container execution and supervision to low-level storage and network attachments.
\containerd supports the Open Container Initiative (OCI) \cite{oci} image and runtime specifications, and provides the substrate for multiple container offerings, such as Docker \cite{docker}, Kubernetes \cite{k8s}, and various cloud offerings \cite{amazon_docker,aci,google_cloud_run}.
Clients interact with \containerd via a client interface, such as \texttt{ctr}, \texttt{nerdctl}, or \texttt{crictl}. \containerd supports both running bare metal containers (\ie those that run directly on the host) and also containers that run within a utility VM (\uvm).

\begin{figure*}
      \centering
      \includegraphics[clip, trim=0.5cm 0.5cm 0.5cm 0.5cm, width=.8\linewidth]{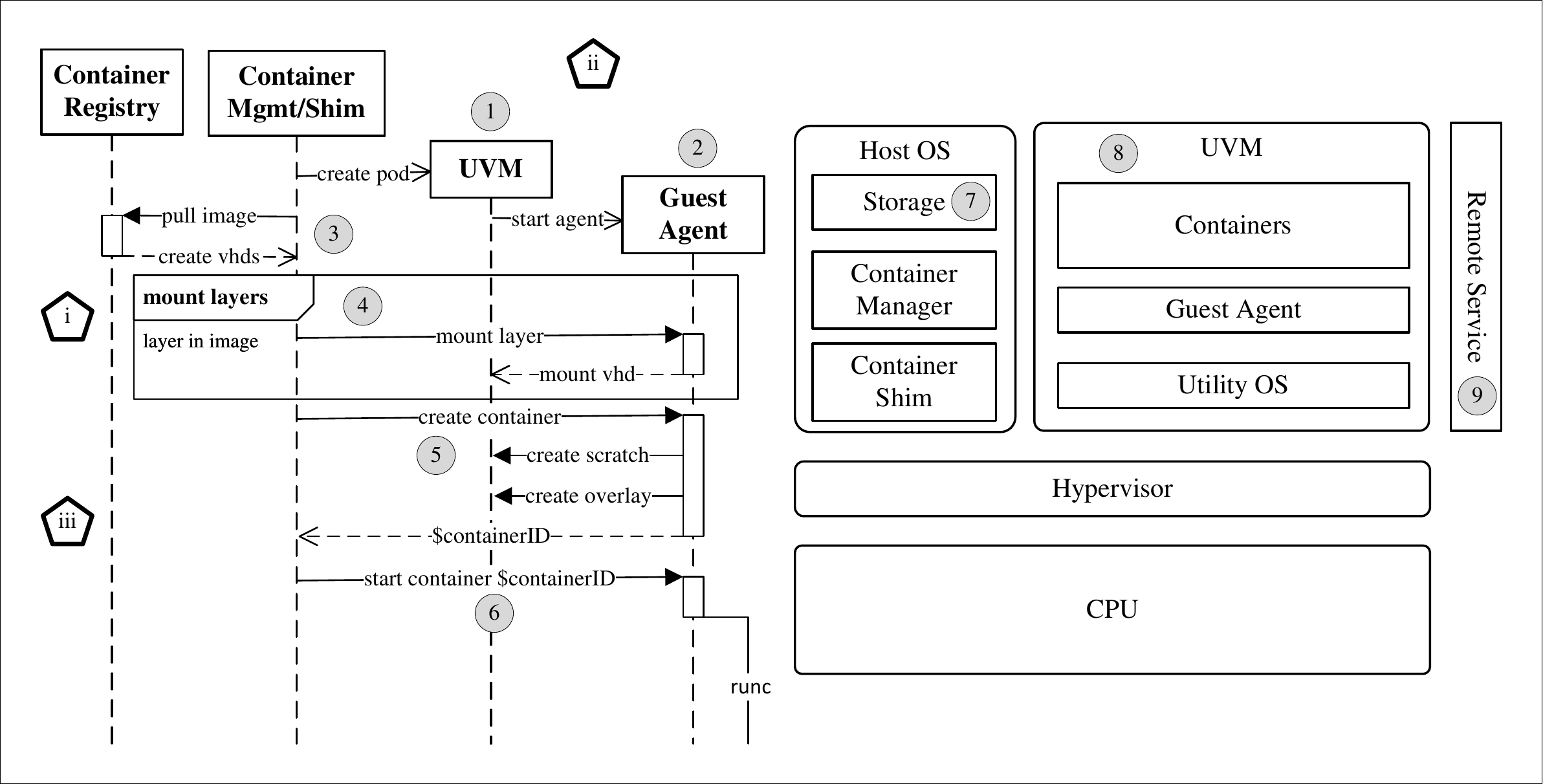}
      \caption{\textbf{Container Flow.} The sequence diagram on the left shows the process that results in a VM-isolated container.
      The pentagons correspond to the workflow steps from the text: (i) pull the image, (ii) launch a pod, (iii) start a container.
      The circles in this figure outline multiple points of attack within this workflow:
      The \hcs may pass a compromised (1) \uvm image or (2) guest agent during \uvm creation.
      The container manager can alter or fabricate malicious layer VHDs (3) and/or mount any combination of layers onto to \uvm (4).
      The \hcs can pass any set of layers to use for creating a container file system (5), as well as any combination of environment variables or commands (6).
      A compromised host OS can tamper with local storage (7), attack the memory of the \uvm (8), or manipulate remote communications (9).
      This list is not comprehensive.
      }
      \label{fig-container-flow}
\end{figure*}

Our work focuses on VM-isolated containers. The \cplat interfaces with a custom \hcs running in the host operating system.
The \hcs interacts with (i) host services to bring up a \uvm required for launching a new pod and (ii) the \gcs running in the \uvm.
The \gcs is responsible for creating containers and spawning a \texttt{runc} instance for starting the container.
In essence, the \hcs-\gcs path allows the \cplat components running in the host operating system to execute containers in isolated guest VMs.
The execution of a VM-isolated container on Linux using \cplat involves three high-level steps  (as seen in Figure \ref{fig-container-flow}):
(i) pull the container's image,
(ii) launch a pod for hosting the container,
(iii) start a container.

\begin{description}
      \item [Image pull] Pulling images entails downloading them from a container registry (unless they are already cached on the machine).
            Once the pull is done, the image is unpacked and the image for each layer of the container is stored as a virtual hard drive.

      \item [Pod launch] Launching a pod entails creating and launching a \uvm along with the
            \gcs. Thereafter, the \hcs interacts with the \gcs to create and start containers.
            At the end of the pod launch, the \hcs creates and starts a sandbox/pause container
            that holds the Linux namespaces for future containers.

      \item[Container start] Starting a container requires that the \gcs mounts the container's root filesystem into the \uvm; the root filesystem comprises the container layers and a writeable scratch layer.
            In doing so, the \hcs (i) attaches each container layer (in the OCI  specification) to the \uvm and (ii) creates and attaches to the \uvm a writeable sandbox virtual hard drive.
            The container layer and sandbox devices are then mounted by the \gcs into the \uvm as an overlay root filesystem.
            Finally, the \gcs creates a runtime bundle that contains the overlay filesystem path and configuration data (compiled using the OCI runtime specification.)
            The runtime bundle is passed to the \texttt{runc} instance, which subsequently starts the container.
\end{description}

%-------------------------------------------------------------------------------
\subsection{Threat Model}
%-------------------------------------------------------------------------------
\label{sec-threat-model}

We consider a strong adversary who controls the entire host system software, including the hypervisor and the host operating system along with all services running within it.
However, we trust the CPU package, including the platform security processor (PSP) and the AMD SEV-SNP implementation, which provides hardware-based isolation of the guest's address space from the system software.
We also trust the firmware running on the PSP and its measurements of the guest VM, including the \gcs.

Such an adversary can:
\begin{itemize}
      \item tamper with the container's OCI runtime specification;
      \item tamper with block devices storing the read-only container image layers and the writeable scratch layer;
      \item tamper with container definitions, including the overlay filesystem (\ie changing the order or injecting rogue layers), adding, altering, or removing environment variables, altering the user command, and the mount sources and destinations from the \uvm.
      \item add, delete, and make arbitrary modifications to network messages, \ie fully control the network.
      \item request execution of arbitrary commands in the \uvm and in individual containers.
      \item request debugging information from the \uvm and running containers, such as access to I/O, the stack, or container properties.
\end{itemize}

These capabilities provide the adversary with the ability to gain access to the address space of the guest operating system.

\paragraph{Out of Scope.} Anything not mentioned here, \eg side-channel attacks, are outside the scope of our threat model.

\begin{figure*}[t]
  \centering
  \includegraphics[width=.7\linewidth]{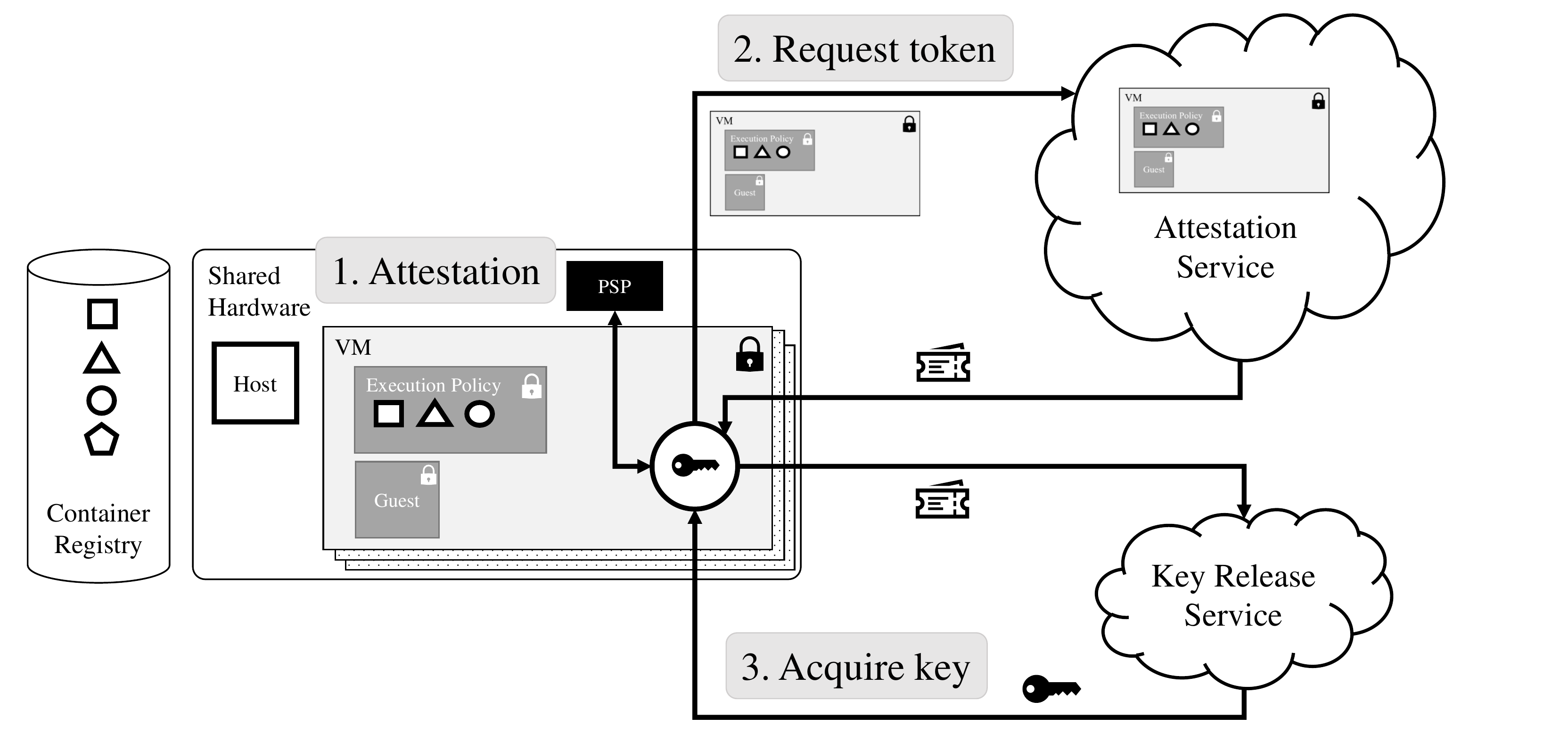}
  \caption{\textbf{Attestation workflow.} Here we present a typical attestation workflow. A container (key in circle) attempts to obtain and decrypt the user's data for use by other containers in the group. The key has been previously provisioned into a key management service
  with a defined key release policy.
    The container within the \uvm requests that the PSP issues an attestation report (1) including an RSA wrapping public key as a runtime claim.
    The report and additional attestation evidence are provided to the attestation service (2), which verifies that it the report is valid and then provides an attestation token that represents platform, init, and runtime claims.
    Finally, the attestation token is provided to the key management service (3) which returns the customer's key to the container wrapped using a RSA public key as long as the token's claims satisfy the key release policy statement.
  }
  \label{fig-attestation}
\end{figure*}

\subsection{Security Guarantees}
\label{sec-guarantees}
Under the threat model presented in Section \ref{sec-threat-model}, we wish to provide strong confidentiality and integrity guarantees for the container and for customer data. The provided security guarantees are based on the following principles:

\vspace{0.1in}
\noindent\textbf{Hardware-based Isolation of the \uvm.} The memory address space and disks of the VM must be protected from the host and other VMs by hardware-level isolation.
\system relies on the SEV-SNP hardware guarantee that the memory address space of the \uvm cannot be accessed by host system software.

\vspace{0.1in}
\noindent\textbf{Integrity-protected Filesystems.} Any block device or blob storage is mounted in the \uvm as an integrity-protected file system. The file system driver enforces integrity checking upon an access to the file system, ensuring that the host system software cannot tamper with the data and container images.
In \system, a container file system is expressed as an ordered sequence of layers, where each layer is mounted as a separate device and then assembled into an overlay filesystem \cite{overlayfs}.
First, \system verifies as each layer is mounted that the dm-verity root hash \cite{dmverity} for the device matches a layer that is enumerated in the policy.
Second, when the \hcs requests the mounting of an overlay filesystem that assembles multiple layer devices, \system verifies that the specific ordering of layers is explicitly laid out in the execution policy for one or more containers.

\vspace{0.1in}
\noindent\textbf{Encrypted Filesystems.} Any block device or blob storage that holds privacy-sensitive data is mounted as an encrypted filesystem.
The filesystem driver decrypts the memory-mapped block upon an access to the filesystem.
The decrypted block is stored in hardware-isolated memory space, ensuring that host system software cannot access the plaintext data.
The writable scratch space of the container is mounted with dm-crypt \cite{dmcrypt} and dm-integrity \cite{dmintegrity}, and this is enforced by the execution policy.
The encryption key for the writeable scratch space is ephemeral and is provisioned initially in hardware-protected memory and erased once the device is mounted.

\vspace{0.1in}
\noindent\textbf{\uvm Measurement.}
The \uvm, its operating system and the \gcs are crytographically measured during initialization by the TEE and this measurement can be requested over a secure channel at any time by user containers.
The AMD SEV-SNP hardware performs the measurement and encodes it in the signed attestation report as discussed in Section~\ref{sec-attestation}.

\vspace{0.1in}
\noindent\textbf{Verifiable Execution Policy.} The user must be provided with a mechanism to verify that the active execution policy (see below in (Section \ref{sec-policy})) in a container group is what they expect it to be.
The execution policy is defined and measured independently by the user and it is then provided to the CaaS deployment system.
The host measures the policy (\eg using SHA-512) and places this measurement in the immutable \textit{host data} of the report as described in Section \ref{sec-attestation}.
The policy itself is passed to the \uvm by the \hcs, where it is measured again to ensure that its measurement matches the one encoded as \textit{host data} in the report.

\vspace{0.1in}
\noindent\textbf{Remote Attestation.} Remote verifiers (\ie tenants, external services, attestation services) need to verify an attestation report so that they can establish trust in a secure communication channel with the container group running within the \uvm.
In particular, remote verifiers need to to verify that the \uvm has booted the expected operating system, the correct \gcs, and further that the \gcs is configured with the expected execution policy.

In \system, the \uvm (including privileged containers) can request an attestation report using the secure channel established between the PSP and the \uvm, as detailed in Section \ref{sec-attestation}.
The requester generates an emphemeral token (\eg TLS public key pair or a sealing/wrapping public key) which is presented as a runtime claim in the report; the token's cryptographic digest is encoded as \textit{report data} in the report.
A remote verifier can then verify that
(i) the report has been signed by a genuine AMD processor using a key rooted to AMD's root certificate authority,
(ii) the \textit{guest launch measurement} and \textit{host data} match the expected VM measurement and the digest of the expected execution policy,
(iii) the \textit{report data} matches the hash digest of the runtime claim presented as additional evidence.

Once the verification completes, the remote verifier that trusts the \uvm (including the guest OS, \gcs and the execution policy) trusts that the \uvm and the container group running with it will not reveal the private keys from which the public tokens have been generated, \eg TLS private key, sealing/wrapping private key.
The remote verifer can utilize the runtime claim accordingly.
For instance,
\begin{itemize}
\item a TLS public key can be used for establishing a TLS connection with the attested container group.
As such, the remote verifier can trust there is no replay or man-in-the-middle attack;
\item a sealing public key can be used to seal (via encryption) a request or response intended only for the attested containers;
\item a wrapping public key can be used by a key management service to wrap and release encryption keys required by the VM's container group for decrypting encrypted remote blob storage.
As such, the remote verifier can trust that only trustworthy and attested container groups can unwrap the encryption keys.
Figure~\ref{fig-attestation} illustrates this process.
\end{itemize}

\subsection{Execution Policy}
\label{sec-policy}
As discussed in our threat model, the \hcs is not trusted as it could be under the control of an attacker.
This implies that any action which the \hcs requests the \gcs undertake inside the \uvm is suspect (see Section \ref{sec-threat-model} for a list of malicious host actions).
Even if the current state of the container group is valid, there is no guarantee that the host will not compromise it in the future, and thus no way for the attestation report to be used as a gate on access to secure customer data.
The attestation report on its own simply records the \uvm OS, the \gcs, and the container runtime versions in use.
It is not able to make any claims about the container group the host will subsequently orchestrate.

For example, the host can start the user container group in a manner which is expected by an attestation service until such time as it acquires some desired secure information, and then load a series of containers which open the container group to a remote code execution attack.
The attestation report, obtained during initialization, cannot protect against this.
Even updating it via providing additional runtime data to the PSP (as described in Section \ref{sec-attestation}) does not help, because the vulnerability is added by the host after the attestation report has been consumed by the external service.

To address this vulnerability, we introduce the concept of an \emph{execution policy}.
Authored by the customer, it describes what actions the \gcs is allowed to take throughout the lifecycle of the container group.
The \gcs is altered to consult this policy before taking any of the actions in Table \ref{tbl-enforcement-points}, providing information to the policy that is used to make decisions.
These actions each have a corresponding \emph{enforcement point} in the execution policy which will either allow or deny the action.
In our implementation the policy is defined using the Rego policy language \cite{rego}.
A sample enforcement point can be seen in Listing \ref{lst-policy-sample}.

\begin{table}
  \begin{tabularx}{\linewidth}{l X}
    \toprule
    \thead{Action}      &
    \thead{Policy Information}                          \\
    \midrule
    Mount a device      &
    device hash, target path                            \\
    \addlinespace[5pt]
    Unmount a device    &
    target path                                         \\
    \addlinespace[5pt]
    Mount overlay       &
    ID, path list, target path                          \\
    \addlinespace[5pt]
    Unmount overlay     &
    target path                                         \\
    \addlinespace[5pt]
    Create container    &
    ID, command, environment, working directory, mounts \\
    \addlinespace[5pt]
    \makecell[lt]{Execute process                       \\
    (in container)}     &
    ID, command, environment, working directory         \\
    \addlinespace[5pt]
    \makecell[lt]{Execute process                       \\
    (in \uvm)}          &
    command, environment, working directory             \\
    \addlinespace[5pt]
    Shutdown container  &
    ID                                                  \\
    \addlinespace[5pt]
    Signal process      &
    ID, signal, command                                 \\
    \addlinespace[5pt]
    Mount host device   &
    target path                                         \\
    \addlinespace[5pt]
    Unmount host device &
    target path                                         \\
    \addlinespace[5pt]
    Mount scratch       &
    target path, encryption flag                        \\
    \addlinespace[5pt]
    Unmount scratch     &
    target path                                         \\
    \addlinespace[5pt]
    Get properties      &
    ---                                                 \\
    \addlinespace[5pt]
    Dump stacks         &
    ---                                                 \\
    \addlinespace[5pt]
    \makecell[lt]{Logging                               \\
    (in the \uvm)}      &
    ---                                                 \\
    \addlinespace[5pt]
    \makecell[lt]{Logging                               \\
    (containers)}       &
    ---                                                 \\
    \bottomrule
  \end{tabularx}
  \caption{\textbf{Policy Actions}. These are the actions we propose to be under the control of the execution policy.
    The list is specific to our implementation, but given standardization around \containerd it should be applicable to most scenarios.
    First we have actions which pertain to the creation of containers.
    By ensuring that any device mounted by the guest has a dm-verity root hash~\cite{dmverity} that is listed in the policy, and that they are combined into overlay filesystems~\cite{overlayfs} in layer orders that coincide with specific containers, we first establish that the container file systems are correct.
    We can then start the container, further ensuring that the environment variables and start command comply with policy and that mounts from the \uvm to the container are as expected (along with other container specific properties).
    Other actions proceed in a similar manner, constraining the control which the \hcs has over the \gcs.
  }
  \label{tbl-enforcement-points}
\end{table}

\begin{lstlisting}[caption={\textbf{Sample enforcement point.} Here, as in our
      implementation, the policy is expressed in Rego \cite{rego}.},
      label=lst-policy-sample]
default mount_device := {"allowed": false}

device_mounted(target) {
  data.metadata.devices[target]
}

mount_device := {"metadata": [addDevice],
                 "allowed": true} {
  not device_mounted(input.target)
  some container in data.policy.containers
  some layer in container.layers
  input.deviceHash == layer
  addDevice := {
      "name": "devices",
      "action": "add",
      "key": input.target,
      "value": input.deviceHash
  }
}
\end{lstlisting}

A novel feature of our implementation is the ability for a policy to manipulate its own metadata state (maintained by the \gcs).
This provides an attested mechanism for the execution policy to build a representation of the state of the container group, allowing for more complex interactions.
For example, in the rule shown in Listing \ref{lst-policy-sample}, the enforcement point for mounting a device creates a metadata entry for the device which will be used to prevent other devices from being mounted to the same target path.

The result of making this small change to the \gcs is that the state space of the container group is bounded by a state machine, in which transitions between states correspond to the actions described above.
Each transition is executed atomically and comes with an enforcement point.

\paragraph{Induction.} The state machine starts as a system that is fully measured and attested, including the execution policy with all its enforcement points, with the root of trust being the PSP hardware ($n=1$).
All possible transitions are described by the execution policy.
Regardless of which ($n$) transitions have been taken after that, each of the actions listed in Table~\ref{tbl-enforcement-points} cannot break integrity or confidentiality without deliberate modification of the respective enforcement point, which would have had to happen before the initial measurement ($n+1$).
Any sequence of such transitions therefore maintains integrity and confidentiality.
Our enforcement points are carefully designed to maintain these properties.
Note that confidentiality is \textit{modulo} acceptance of the \uvm and the execution policy.
That is, the end-user must verify that the attestation report they receive from \system is bound to a \uvm and an execution policy that uses the end-user's data in a manner they accept.

%-------------------------------------------------------------------------------
\section{Evaluation}
%-------------------------------------------------------------------------------

We used benchmarking tools to evaluate several typical containerized workloads for reductions in computation throughput, network throughput, and database transaction rates to ensure that \system does not introduce significant computational overhead.
In all cases we demonstrate that \system provides confidentiality for containerized workloads with minimal costs to performance (typically less than 1\% additional overhead over running in an enclave).

Each benchmarking experiment is conducted using two machines:
(1) a DELL PowerEdge R7515 with an AMD EPYC 7543P 32-Core Processor and 128GB of memory for hosting the container runtime and
(2) a benchmarking client (to avoid impact of any benchmarking software upon the evaluation) with the same configuration connected to (1) on the same subnet via 10GBit Ethernet.
(1) is running Windows Server 2022 Datacenter (22H2) and an offline version of the Azure Container Instances \cplat (\ie \texttt{containerd}, \texttt{cri}, and \texttt{hcsshim}).
The \uvm runs a patched version of 5.15 Linux which includes AMD and Microsoft patches to provide AMD SEV-SNP enlightenment.
(2) is running Ubuntu 20.04 with Linux kernel 5.15.

\vspace{-5pt}
\subsection{\nginx}
\begin{figure}
    \includegraphics[width=\linewidth]{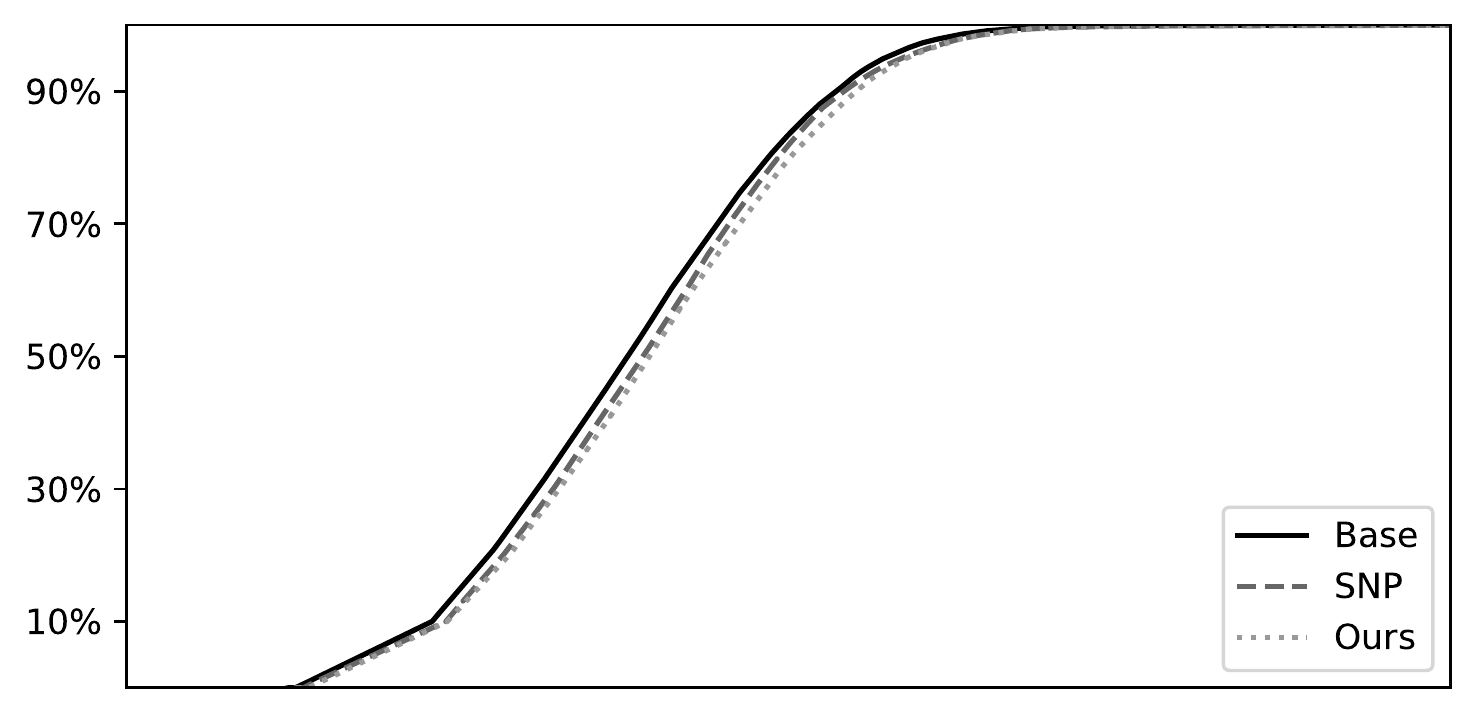}
    \includegraphics[width=\linewidth]{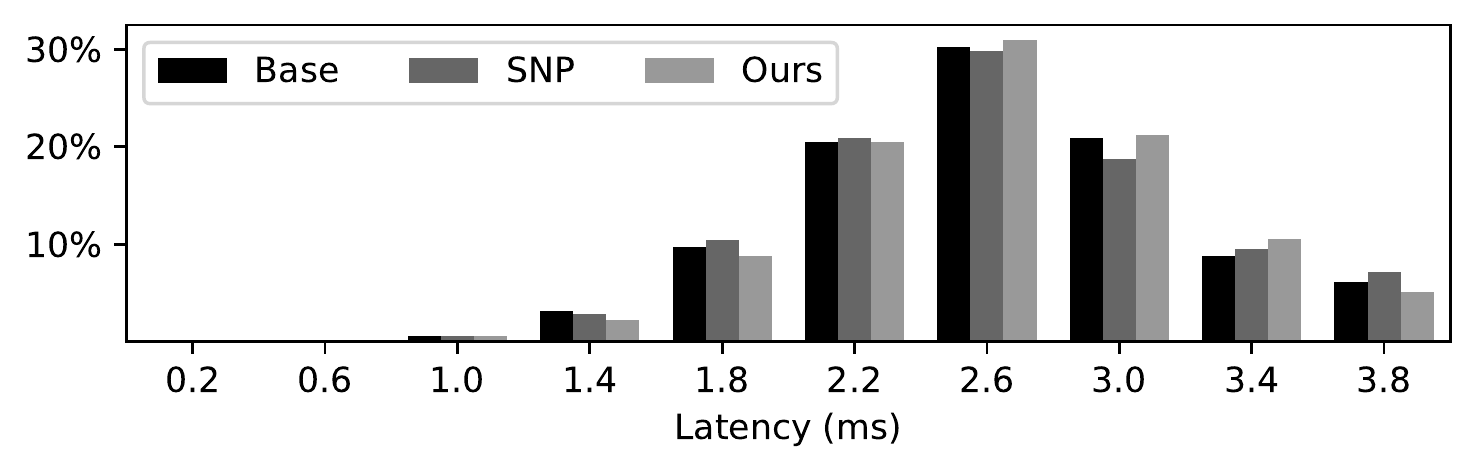}
    \caption{\textbf{\nginx Results}. Here we see results from our benchmarking of \nginx.
        The top plot shows the median latency curves (left and up is better).
        The bottom plot shows the latency histograms.
        We gather latency samples from 20 experiments in which we simulate 100 clients making 200 requests a second, for 60 seconds.
        The overhead of running inside the enclave introduces a noticeable increase in latency in SEV-SNP over Base.
        The addition in \system of the execution policy, however, results in a less than 1\% increase in latency over SEV-SNP.}
    \label{fig-nginx}
\end{figure}

Web services are a common use case for containerization, and so we benchmark the popular \nginx webserver using the \texttt{wrk2} \cite{wrk2} benchmarking tool.
Each test is run for 60 seconds on 10 threads, simulating 100 concurrent users making 200 requests per second (for a total of 12000 requests per test).
We repeat the tests 20 times for each of three configurations:
\begin{description}
    \item[Base] A baseline \nginx container running outside the SEV-SNP enclave,
    \item[SEV-SNP] The same container running within the SEV-SNP enclave,
    \item[\system] The same container again within the enclave and with an attested execution policy,
\end{description}
and measure the latency. The results are shown in Figure \ref{fig-nginx}.
The  median curves are computed over all experiments per configuration, and the histograms are composed of all latency samples which were gathered.
We observe an increase in latency by the introduction of SEV-SNP, as expected, and also a very minor increase in latency when adding the execution policy.
However, it is worth noting that these effects are only reliably observed in aggregate, \ie all median curves are within the first quartiles of each other.

\vspace{-5pt}
\subsection{\redis}
The in-memory key/value database \redis provides another useful benchmark for containerized compute.
It supports a diverse set of data structures, such as hashtables, sets, and lists.
We perform our benchmarking using the provided \texttt{redis-benchmark} with 25 parallel clients and a subset of tests, the results of which can be seen in Table \ref{tbl-redis}.
Looking at the geometric mean over all actions, we see a performance overhead of 18\% added by operating within the AMD SEV-SNP enclave, and a further 1\% when using \system.
The performance overhead is attributed to increased TLB pressure arising from large working sets which exhibit poor temporal reuse in TLBs and trigger page table walks.
In SEV-SNP-enabled systems, table page table walks incur additional metadata checks (in the Reverse Page Table) to ensure that the page is indeed owned by the VM.

\begin{table}
    \centering
    \begin{tabularx}{\linewidth}{X c c c c c c}
        \toprule
        \thead{Setup}                                 &
        \rotatebox{90}{\thead{\texttt{PING\_INLINE}}} &
        \rotatebox{90}{\thead{\texttt{PING\_BULK}}}   &
        \rotatebox{90}{\thead{\texttt{SET}}}          &
        \rotatebox{90}{\thead{\texttt{GET}}}          &
        \rotatebox{90}{\thead{\texttt{INCR}}}         &
        \rotatebox{90}{\thead{\texttt{LPUSH}}}                                                                                                                                                                                                        \\
        \midrule
        Base                                          & {\small 68{\scriptsize ±2.3}} & {\small 69{\scriptsize ±2.3}} & {\small 69{\scriptsize ±1.8}} & {\small 69{\scriptsize ±1.9}} & {\small 69{\scriptsize ±2.1}} & {\small 68{\scriptsize ±2.0}} \\
        SNP                                           & {\small 57{\scriptsize ±3.2}} & {\small 58{\scriptsize ±4.7}} & {\small 57{\scriptsize ±3.1}} & {\small 57{\scriptsize ±4.2}} & {\small 57{\scriptsize ±3.7}} & {\small 55{\scriptsize ±4.4}} \\
        Ours                                       & {\small 56{\scriptsize ±3.3}} & {\small 57{\scriptsize ±3.4}} & {\small 56{\scriptsize ±4.1}} & {\small 56{\scriptsize ±5.0}} & {\small 54{\scriptsize ±4.3}} & {\small 54{\scriptsize ±4.5}} \\
    \end{tabularx}
    \begin{tabularx}{\linewidth}{c c c c c c X}
        \toprule
        \rotatebox{90}{\thead{\texttt{RPUSH}}} &
        \rotatebox{90}{\thead{\texttt{LPOP}}}  &
        \rotatebox{90}{\thead{\texttt{RPOP}}}  &
        \rotatebox{90}{\thead{\texttt{SADD}}}  &
        \rotatebox{90}{\thead{\texttt{HSET}}}  &
        \rotatebox{90}{\thead{\texttt{SPOP}}}  &
        \rotatebox{90}{\thead{GeoMean}} \\
        \midrule
        {\small 69{\scriptsize ±2.3}} & {\small 69{\scriptsize ±2.3}} & {\small 68{\scriptsize ±2.1}} & {\small 69{\scriptsize ±1.7}} & {\small 67{\scriptsize ±1.4}} & {\small 69{\scriptsize ±2.1}} & {\small 68}{\scriptsize ±1.0}\\
        {\small 56{\scriptsize ±4.1}} & {\small 54{\scriptsize ±4.2}} & {\small 55{\scriptsize ±4.2}} & {\small 58{\scriptsize ±3.5}} & {\small 54{\scriptsize ±4.3}} & {\small 57{\scriptsize ±3.9}} & {\small 56}{\scriptsize ±1.1}\\
        {\small 54{\scriptsize ±3.8}} & {\small 52{\scriptsize ±3.9}} & {\small 54{\scriptsize ±4.3}} & {\small 56{\scriptsize ±4.0}} & {\small 56{\scriptsize ±4.5}} & {\small 58{\scriptsize ±4.4}} & {\small 55}{\scriptsize ±1.1}\\        \bottomrule
    \end{tabularx}
    \caption{\textbf{\redis results.}. In this table we show the comparative request rates for different tests from the standard \texttt{redis-benchmark} tool broken out by Base (container run without SEV-SNP), SNP (with SEV-SNP), and \system (SEV-SNP + execution policy).
        Values are in thousands of requests per second, higher is better.
        While both SEV-SNP and \system exhibit a decrease in performance over Base as a result of the overheads introduced by running within the enclave, there is little difference between them.}
    \label{tbl-redis}
\end{table}

\subsection{SPEC2017}
\begin{figure*}
    \centering
    \includegraphics[width=.8\linewidth]{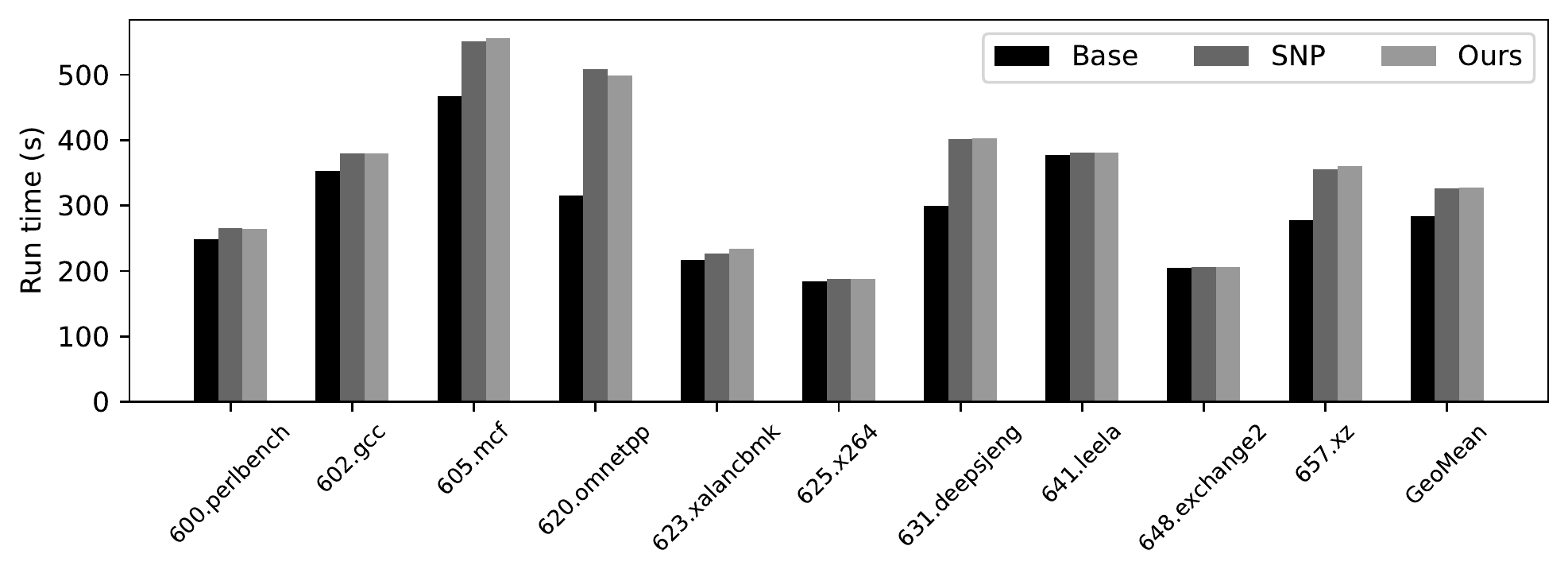}
    \caption{\textbf{SPEC2017 results}. This plot shows the per-benchmark runtimes (in seconds) for SPEC2017 broken out by Base (container run without SEV-SNP), SNP (with SEV-SNP), and \system (SEV-SNP + execution policy).
        Lower is better.
        \texttt{631.deepsjeng}, the most memory-intensive benchmark in SPECint introduces the second-highest overhead.
        \texttt{620.omnetpp}, the most IO-intensive benchmark (due to large test inputs) introduces the highest overhead.
        Discussions of these outliers can be found in the text.}
    \label{fig-spec}
\end{figure*}

We also evaluate \system by measuring the computation performance overhead using the SPEC2017 \texttt{intspeed} benchmarks \cite{spec2017}.
The benchmark programs are compiled and run on the bare metal hardware.
When containerized, they are provided with 32 cores and 32 GB of memory.
As can be seen in Table \ref{tbl-spec} AMD SEV-SNP adds a performance overhead of 13\% on average, and \system adds less than 1\% on top of this.
By looking at the individual benchmarks in Figure \ref{fig-spec}, SEV-SNP introduces a wide range (1-38\%) of performance overhead in SPECint benchmarks.
The overheads are down to (i) the increased TLB pressure (further exacerbated in SEV-SNP setups as discussed in \texttt{redis} benchmark);
\texttt{631.deepsjeng}, the most memory-intensive benchmark in SPECint introduces the second-highest overhead and
(ii) the additional overhead for accessing the encrypted scratch space; \texttt{620.omnetpp}, the most IO-intensive benchmark (due to large test inputs) introduces the highest overhead (38\%).

\begin{table}
    \centering
    \begin{tabularx}{.6\linewidth}{X c c c}        \toprule
        \thead{Setup}      &
        \thead{Base Ratio} &
        \thead{- (Base) }                  \\
        \midrule
        Base               & 8.43 & 0.0\%  \\
        SNP                & 7.33 & 13.1\% \\
        Ours           & 7.30 & 13.4\% \\
        \bottomrule
    \end{tabularx}
    \caption{\textbf{SPEC2017 results.} We run the SPEC2017 \texttt{intspeed} benchmarks in four configurations:
        on the bare metal host (Bare),
        in a container (Base),
        in a container in a hardware-isolated VM (SNP),
        and Ours (\ie using \system).
        The reported values are the base ratio from reportable benchmark runs (see \cite{spec2017} for details on the \texttt{reportable} flag).
        Higher is better.}
    \label{tbl-spec}
\end{table}

\subsection{NVidia Triton Inference Server}
Finally, we evaluate \system by running a machine learning (ML) inference workload based on NVidia's Triton Inference Server \cite{triton}. Models (trained offline) and their parameters are used by the server to serve requests via a REST API.

The confidential ML inference server is deployed via a container group that comprises two containers: an (unmodified) Triton inference container, and a sidecar container that mounts an encrypted remote blob (holding the ML model) using  dm-crypt and dm-integrity.
(The sidecar container also implements the attestation workflow described in Figure~\ref {fig-attestation} to release the encryption key.)
The filesystem (and the contained ML model) is made available to the Triton inference container.

\begin{figure}
    \includegraphics[width=\linewidth]{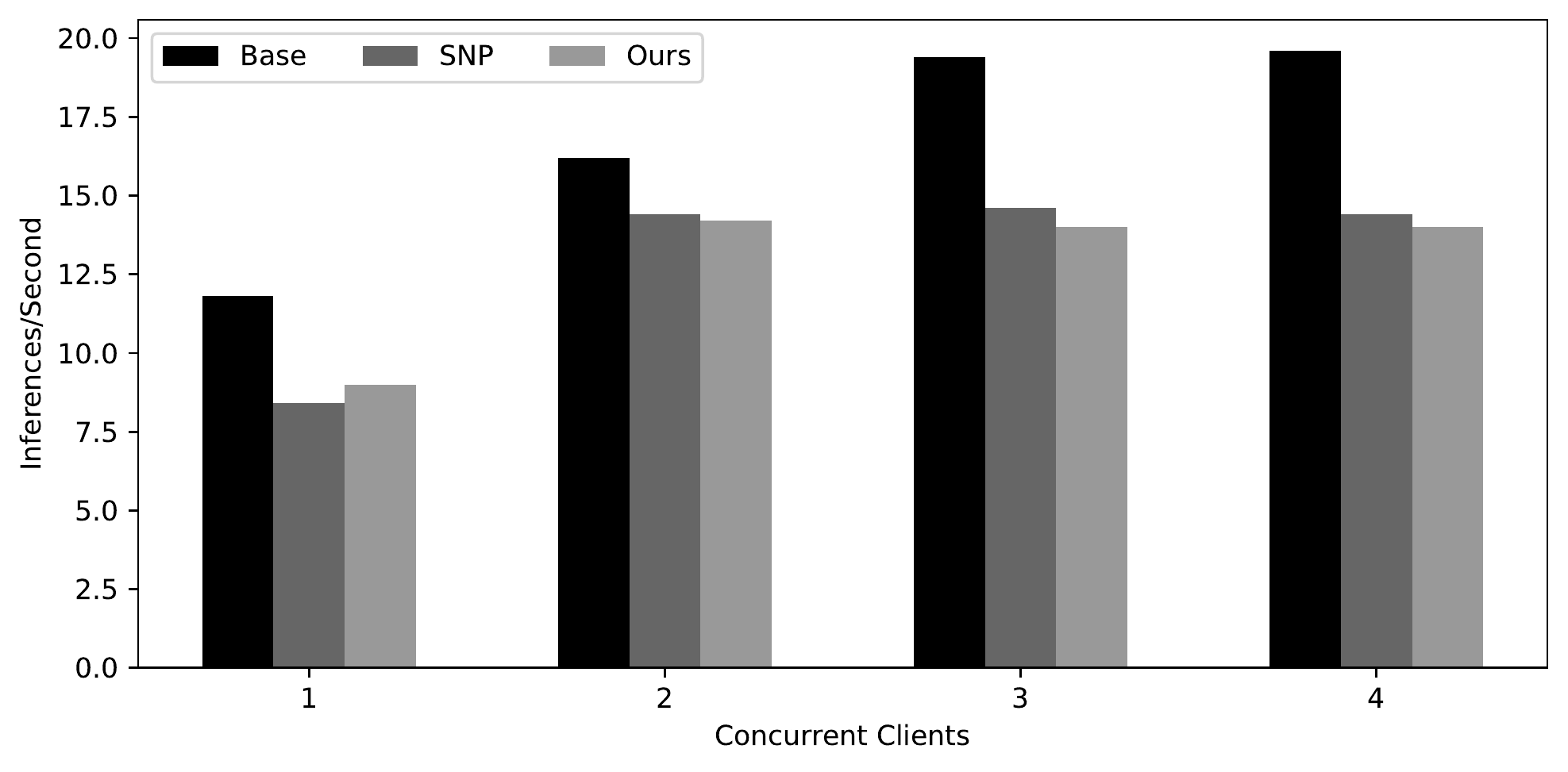}
    \caption{\textbf{NVidia Triton Results}. This plot shows the inference rate for the Base (container run without SEV-SNP), SNP (with SEV-SNP), and \system (SEV-SNP + execution policy) configurations. The median values gathered over three experiments are shown for each number of concurrent clients.}
    \label{fig-triton}
\end{figure}

We evaluate the inference servers using NVidia's \texttt{perf-analyzer} system, allowing us to measure the overhead introduced by SEV-SNP and \system, as shown in Figure \ref{fig-triton}.
For each of the three configurations (Base, SNP, and \system) we run four different experiments with 1 to 4 concurrent clients making continuous inference requests.
In Figure \ref{fig-triton} we report the median throughput for these experiments over 3 trials and observce a performance overhead of 26\% when running in the AMD SEV-SNP enclave.
As before, the additional overhead from \system is ~1\%.
The overheads share the same root cause in the increased TLB pressure as was previous described in the \texttt{redis} benchmark.

%-------------------------------------------------------------------------------
\section{Related Work}
%-------------------------------------------------------------------------------

A number of TEE container runtimes have been proposed to enable running applications or containers on Intel SGX \cite{arnautov16scone, baumann14haven, priebe20sgxlkl, shinde17panoply, tsai17graphenesgx}.
A common feature across these proposals is the use of a library OS running in-enclave.
By design, a library OS provides a subset of OS features, and so can only run some containers without modification.
In addition, the application's network interface, the interface between the library OS, and the actual out-of-enclave OS are security boundaries.
This has a performance impact, as the library OS must provide a secure communication mechanism to the out-of-enclave OS and validate all data that crosses the boundary.
In contrast, \system provides an actual OS in-enclave, can run unmodified containers, and only the application's network interface is a security boundary.
Additionally, \system provides the inductive proof of all future states via the attested execution policy.

Hecate \cite{ge2022hecate} uses AMD VM privilege levels (VMPLs) to run a nested hypervisor and a guest OS within the same AMD SEV-SNP isolated VM.
This allows an unmodified guest OS to be run as a confidential VM, modulo kernel code integrity.
However, Hecate does not address attestation, file system integrity and confidentiality, or execution policy.
In addition, Hecate allows guest OS administrators full access to the VM.

Brasser \etal have concurrently proposed TCX, a collection of trusted container extensions for running containers securely within hardware-isolated utility VMs on AMD SEV processors \cite{brasser22tcx}.
TCX relies (a) on a root VM (akin to the SGX quoting enclave) for bootstrapping the utility VM, thus increasing the TCB, and
(b) on a secure channel established between the container owner and the utility VM for preventing untrusted entities from submitting container commands to the VM.
The latter design choice does not support the CaaS deployment model, wherein the container owner and cloud service provider personas are different, thus requiring that the CSP submit container commands to the utility VM.
In addition, the container owner is in the TCB.
The end-user of the confidential container (\eg a bank customer, a patient submitting medical data to their doctor) has no attestation over what container commands have been or will be run.

SEVGuard explores running user-mode applications on guest SEV-protected VMs without a guest-side kernel component \cite{palutke19sevguard}.
SEVGuard relies on an existing kernel virtualization API for interaction with host kernel features and provides support for calling shared libraries on the host.
While SEVGuard offers a low TCB, it is vulnerable to attacks on the kernel virtualization API and shared libraries and does not provide secure persistent storage.

%-------------------------------------------------------------------------------
\section{Future Work}
%-------------------------------------------------------------------------------

While \system provides a solid foundation for confidential containers, there are some limitations to this technique which invite future investigation.

\subsection{Trusted Computing Base}
\system reduces the TCB by removing the need to trust the host, the hypervisor, and the CSP, but it could be smaller.
In particular, by trusting the \uvm we necessarily import the \uvm OS (\eg Linux) into the TCB, as well as the standard libraries needed to implement other elements like the \gcs.
However, much of that code is entirely vestigial in the context of providing a container runtime.
One potential line of inquiry would be to explore ways of reducing this aspect of the \uvm to the smallest possible kernel and the barest necessities needed by the \gcs, \texttt{runc} and other tools to further reduce the attack surface.

\subsection{Policy Flexibility}
One downside of having an execution policy which is measured during initialization and then subsequently used for attestation-based security operations is that the release policies will necessarily be tied to a fixed version of the execution policy.
If container images need to change, \eg due to necessary security updates upon discovery of a vulnerability, it requires not only an update to the execution policy but also an update to all release policies as well.
In many scenarios this is a desirable property, but users may want to choose to loosen how the policy defines which actions it allows.
A promising area of future research would be to find a manner in which to provide this flexibility without sacrificing the post-verifiable inductive proof over the state of the container group which \system provides.

\subsection{Writeable Filesystems Freshness}
\system relies on dm-integrity for block-level integrity of writeable filesystems.
While dm-integrity provides integrity protection based on authentication tags, the latter are vulnerable to replay attacks and do not provide any freshness guarantees.
Freshness could be provided using update-able integrity trees (\eg Merkle Trees) but at a huge latency and bandwidth overhead when a data block (\ie leaf block in the tree) is updated due to a chain of updates to all intermediate blocks lying with the root-leaf path.
A promising area of future research would be to explore security-performance trade-offs for writeable filesystems with freshness guarantees.
%-------------------------------------------------------------------------------
\section{Conclusion}
%-------------------------------------------------------------------------------

In this paper we have introduced \system, a novel method for providing confidential computation for containerized workflows via the introduction of an attested execution policy.
Further, we have demonstrated that \system adds less than 1\% additional performance overhead beyond that added by the underlying TEE (\ie AMD SEV-SNP).
We also outline how the security properties of the system provide an inductive proof over the future state of the container group rooted in the attestation report.
This provides the ability (via remote attestation) for external third-parties to securely communicate with containers, enabling a wide range of containerized workflows which require confidential access to secure data.

%-------------------------------------------------------------------------------
\section*{Availability}
%-------------------------------------------------------------------------------

The open source implementation of \system is available as part of the \texttt{hcsshim} system (\url{https://github.com/microsoft/hcsshim/tree/main/pkg/securitypolicy}) and is the technology which enables Confidential Azure Container Instances.

%-------------------------------------------------------------------------------
\section*{Acknowledgements}
%-------------------------------------------------------------------------------
%
Thanks to Istvan Haller for help with the SPEC2017 benchmark and
helpful conversations.

\bibliography{main}{}
\bibliographystyle{plain}

%%%%%%%%%%%%%%%%%%%%%%%%%%%%%%%%%%%%%%%%%%%%%%%%%%%%%%%%%%%%%%%%%%%%%%%%%%%%%%%%
\end{document}